\documentclass[useAMS,usenatbib]{mn2e}
\usepackage[usenames]{color}
\usepackage{ifthen}
\usepackage{natbib}
\usepackage{epsfig}
\usepackage{amssymb}
\usepackage{amsmath}
\usepackage[all]{xy}

\newcommand{\beq}{\begin{equation}}
\newcommand{\eeq}{\end{equation}}

\newcommand{\role}{{role}}
\newcommand{\We}{{\rm We}}
\newcommand{\re}{{{\rm Re}}}
\newcommand{\aSS}{{\alpha_{{}_{\rm SS}}}}

\newcommand{\aLP}{{\alpha_{{}_{\rm L\!BP}}}}
\newcommand{\aW}{{\alpha_{{}_{\rm W01}}}}
\newcommand{\bW}{{\beta_{{}_{\rm W01}}}}
\newcommand{\myM}{{\cal M}}
\newcommand{\bs}{\boldsymbol}

\newcommand{\simgeq}{\; \raisebox{-0.4ex}{\tiny$\stackrel{{\textstyle>}}{\sim}$}\;}
\newcommand{\simleq}{\; \raisebox{-0.4ex}{\tiny$\stackrel{{\textstyle<}}{\sim}$}\;}

\title[Three routes to collimation by the MRI]{Three routes to jet collimation by the Balbus-Hawley magnetorotational instability}
\author[P. T. Williams]{Peter T. Williams$^{1,2}$\thanks{email: petwil@gmail.com, ptw@lanl.gov}\thanks{Guest Scientist, Applied Physics Division, Los Alamos National Laboratory, Los Alamos, NM 87545, USA}\\
$^{1}$405 14$^{\rm th}$ Street, Suite 1207, Oakland, CA 94612, USA\\
$^{2}$Department of Astronomy, City College of San Francisco, San Francisco, CA 94112, USA\\}
\begin{document}

\date{Accepted 2005 May 3. Received 2005 May 2; in original form 2004 December 21}

\maketitle

\begin{abstract}
Three completely different lines of work have recently led to the conclusion
that the magnetorotational instability (MRI) may create a hoop-stress that collimates jets.
One argument  %, which forms the primary focus of our work here,
% is based upon consideration that turbulence created by the MRI is more nearly
 is based upon consideration that magnetohydrodynamic turbulence, in general, 
and turbulence driven by the MRI, in particular, is more nearly
viscoelastic than it is viscous. 
Another argument is based upon the dispersion relation for
the MRI in the context of 1D simulations of core collapse.
 Yet a third argument rests in the results of direct numerical magnetohydrodynamic simulations
of collapsars and thick accretion flows.
I elaborate on my previous work regarding the first argument above and I briefly discuss how these three sets of results are all related. 
I also discuss the different {\role}s played
by the magnetic tension 
and the 
magnetic pressure 
within the context of this work.
I point out that this leads to consideration of the normal stress difference between the hoop stress and the radial stress,
in preference to a focus on just
the hoop stress itself.
Additionally, I argue that simulations of thick accretion flows and collapsars are
 not self-consistent if they include a phenomenological model for
an MRI-induced viscous stress but disregard these
other MRI-induced stress components. I comment briefly on the {\em RHESSI} observation 
of
polarization in the gamma-ray burst GRB0212206. I argue that this polarization is consistent with a tangled field, and
does not require a large-scale organized field. Finally I suggest 
that the role of magnetic fields in creating jets, as described here, 
should be understood not to work within the confines of magnetocentrifugal
models of jets, but rather as an alternative to them.

\end{abstract}

\section{INTRODUCTION}
It has recently been suggested by a number of authors that a magnetic field self-consistently generated by 
the magnetorotational instability (MRI) can collimate jets.
This has been pointed out, independently, following three completely separate lines of work.
Two are based in simple analytical arguments, and a third rests in the interpretation of numerical simulation.

These arguments have been placed in a variety of contexts, including 
supernovae and collapsars, active galactic
nuclei, and protostellar jets. Certain aspects of this previous work are admittedly contextually
specific. None the less I argue that MRI-generated hoop-stresses may be a nearly
universal aspect of astrophysical jet creation and collimation in accretion or collapse scenarios. I explore this possibility
below, where I highlight similarities in these three lines of reasoning. However, I focus primarily below on expounding on
my previous work, as this has not appeared to date in peer-reviewed form.

In Williams~(2001) (hereafter W01) I pointed out that, in contrast with purely hydrodynamic turbulence, magnetohydrodynamic (MHD)
turbulence in an azimuthally shearing environment creates a hoop-stress,\footnote{More precisely, I pointed out that it creates a positive difference between the
azimuthal hoop stress and the radial stress.} 
 in analogy to the hoop-stress in viscoelastic
media undergoing azimuthal shear.
I estimated the magnitude of this stress in a thick accretion disk (or flow) using simple dimensional arguments,
and I suggested that this hoop-stress could collimate jets. Although I discussed stresses in the general terms
of MHD turbulence drawing its energy from azimuthal shear, 
I specifically pointed out in W01 (viz, p.~2, paragraph 1) that,
as should be expected in an accretion scenario, this
includes the Balbus-Hawley instability, {\em i.e.}\/\ the MRI. (It is not clear what other
MHD instabilities exist that may draw their energy from the differential shear in the context of accretion flows.)
This specific point regarding the {\role} of the MRI was later emphasized in \citet{Will2003} (hereafter W03). 
The context of W01 was a generic thick, steady accretion
disk or flow, and in W03 I discussed protostellar jets, although most of the latter
argument is broadly applicable to other jet phenomena.

It was suggested by \cite{AWML2003}, see
also \cite{AkiWhe2002}, that a toroidal field generated by the MRI could contribute to jet creation in 
core-collapse supernovae.
This argument was based upon the dispersion relation for the MRI.
The MRI is a weak-field instability in the sense that the MRI is unstable so long as the magnetic field is less 
than some critical field strength. In addition, the growth rate of the MRI is comparable to the local angular velocity $\Omega$,
which enables several $e$-foldings of the field strength in the course of core-collapse. It was thus hypothesized that the MRI would
amplify a small seed field until it grew to equal the critical field strength, and it was argued that the resultant magnetic field would
contribute to jet creation. This tentative conclusion was bolstered by 1D core-collapse
simulations in which the MRI was treated heuristically through its dispersion relation.

Collimation by an MRI-generated field has also been argued to be happening in several recent direct numerical simulations (DNS) of both
transient and quasi-steady accretion flows. This is particularly
intriguing because DNS simulations, by definition, make no assumptions about the MRI or the ensuing turbulence.
The first simulations in 3D seem to show collimation in the form of a funnel surrounded by high magnetic pressure regions  (\citealt{HBS2001}, \citealt{HawBal2002}).
 This is quite
different from the collimation by hoop-stress that I have argued. Meridional plane ($2.5\,{\rm D}$) simulations
of accretion flows in a variety of contexts seem to show collimation by stresses that the authors argue are generated by the MRI (\citealt{Kud2002}, \citealt{ProBeg2003}, \citealt{PMAB2003}). 
Further numerical simulation and study should help clarify the {\role} of the MRI in this collimation.

\section{COLLIMATION BY TURBULENCE}
\subsection{A preliminary zeroth route}
Tables for the components of the Reynolds stress and turbulent Maxwell stress tensors in shearing-sheet
 simulations of the MRI have been provided by \cite{Br.et95}, \cite{HGB1995} and \cite{HGB1996} (hereafter BNST95, HGB95 and HGB96).
The stresses are due largely to a tangled magnetic field, and the streamwise stress, corresponding to a toroidal hoop stress, is positive (indicating tension) and larger than the (off-diagonal) viscous stress;
in contrast, the radial normal stress is negative (indicating pressure).
Tables for the force due to a stress tensor ({\i.e.} its divergence) can be found in a variety of standard reference sources: 
 A positive hoop-stress 
$W_{\theta \theta}$ and a negative radial stress $W_{RR}$ result in a cylindrically radially-inwards (collimating) force
density $F^{(2)}_R = - (W_{\theta \theta} - W_{RR}) / R$. 
This leads immediately to a  zeroth conclusion that the MRI
can create a collimating force\footnote{
When discussing collimation here and in my previous work, it should be clear that I am not discussing far-downstream collimation, such as in traditional magnetocentrifugal mechanisms.
Rather, I mean collimation in the jet-formation region, including the equatorial plane of accretion. 
}, and that it can do so through a turbulent, tangled field, 
by direct inspection of the results of BNST95 and HGB96. 
The primary serious potential objection to this is that the magnitude of the collimating force
may not be significant, since the MRI saturates at a rather high plasma $\beta_p$ in these simulations.
There remain several additional questions, such as 
how magnetic pressure affects this collimation. In particular note that the radial stress $W_{RR}$ is due largely
to magnetic pressure, and the gradient term $\partial_R{W_{RR}}$
in the divergence of the stress [$F^{(1)}$, infra]
in a real accretion flow  can not be determined from local shearing-sheet simulations. 
The remainder of this section discusses these and other observations and questions regarding the MRI in greater detail,
with the help of some simple turbulence modeling considerations, following and extending the discussion in
W01 and W03.

\subsection{Some simple turbulence modeling considerations}
\subsubsection{Introduction and notation}
By viscosity, unless otherwise noted, I mean shear viscosity and not bulk viscosity.
For the purposes of this paper, by viscous stress I mean that part of a stress tensor $w$, turbulent or otherwise,
which may be written in the explicit tensor form
\beq
w{{}_{\rm visc}} = \nu \left[ \nabla v + (\nabla v)^{\rm T} - {2 \over 3} (\nabla \cdot v) I \right],
\label{eq:viscdef}
\eeq
where $v$ is the velocity and $I$ is the identity.
The only cases considered here are rectilinear shear, such as $\vec v \propto y \hat x$ in Cartesian coordinates, and
azimuthal shear such as in an accretion disk with $R\Omega \gg v_R$. In these two cases, the viscous stress tensor above
reduces to a single, off-diagonal component in the respective coordinate systems, namely $w_{xy}$ and $w{{}_{R\theta}}$.
It is assumed that this stress component is due entirely to viscosity, so that the value of the kinematic viscosity $\nu$,
be it a molecular or a turbulent viscosity, is given by $\nu = w / \gamma$ where $\gamma$ is the shear rate and $w$ is the appropriate
stress component above, and care is taken so that the sign of $\nu$ is consistent with the tensor equation ({\ref{eq:viscdef}) above.

My approach here as elsewhere is to ignore the mean field
entirely, as counterpoint to analyses that ignore the turbulent field entirely.
Thus unless noted otherwise I assume throughout the remainder of Section $2$ that
 the mean field $\left<{\boldsymbol B}\right>$ is identically zero, and the
magnetic field exerts a dynamical influence entirely through a tangled field ${\bs B}'$. I denote the averaged Faraday tension as $4\pi M_{ij} \equiv \left<B'_i B'_j \right>$,
and the turbulent Maxwell stress ${\bs {\myM}}$ is given by ${\myM}_{ij} = M_{ij} - (1/2)M_{kk}\delta_{ij}$. 
The Reynolds stress is ${R}_{ij} \equiv \left< \rho v'_i v'_j \right>$, and the full 
turbulent stress tensor ${\bs W}$ is given by $W_{ij} = -{R}_{ij} + {\myM}_{ij}$.

Note that it is commonplace to refer to the funicular turbulent stress $M_{ij}$ as the
Maxwell stress. This is strictly speaking incorrect; one must not neglect the contribution of the
magnetic pressure term to the turbulent stress tensor. Inclusion of the turbulent magnetic pressure term
does not affect the viscous term nor the normal stress difference, but it is none the less important
in a full consideration of the dynamics of how MHD turbulence collimates jets.

Also for future reference, distinguish here 
between the angular velocity or Coriolis rate $\Omega$,
the shear rate $2 A = - R \partial_R \Omega$ (here $A$ is the first galactic Oort constant), and
the vorticity $(1/R)\partial_R\!\left(R^2 \Omega\right) = 2(\Omega - A)$.
In the case $\Omega \propto R^{-q}$, these may be written $\Omega$, $q\Omega$, and $(2-q)\Omega$ respectively.

\subsubsection{Failure of $\alpha$-prescription and existence of normal stress difference}
The original way \citep{ShaSun1973}
 to write the Shakura-Sunyaev viscosity prescription, under the assumption that $v_R \ll v_\theta$,
 is that the viscous component of the turbulent stress
tensor is proportional to the density times the sound speed squared, $W_{R\theta} = -\aSS \rho c_s^2$.
This represents only one of the six
independent components of the turbulent stress. Through vertical integration, for a quasi-2D ($R$--$\theta$) disk, these
six reduce to three, namely the aforementioned viscous stress and the on-diagonal stresses $W_{RR}$ and $W_{\theta \theta}$.
These on-diagonal components of the turbulent stress are usually ignored.
If anything, it is typically assumed that they consist either simply of the
so-called turbulent pressure, $W_{RR} = W_{\theta \theta} = -P_{\rm turb}$, or are due to a turbulent bulk viscosity.
Either assumption predicts that the on-diagonal stresses are equal to one another. 
Particularly in the case of MHD turbulence, this may be highly inaccurate. %assumption.

On-diagonal stress may be dynamically significant
even when the turbulent pressure and bulk viscosity are not, because of normal stress differences.
It is therefore useful to distinguish between what I call the primitive $\alpha$--viscosity prescription above
where only the viscous stress $W_{R\theta}$ is modeled,
and the extended $\alpha$--viscosity prescription in which the full turbulent stress tensor is modeled as a purely viscous stress,
plus perhaps a pressure term.

The shearing-sheet simulations of the MRI mentioned above 
clearly show that: (i) the stress is dominated by the turbulent Maxwell stress,
rather than the Reynolds stress, and (ii) as pointed out above, the cross-stream stress and
the streamwise stress are of opposite sign, and both are
 actually larger in magnitude than the viscous stress, creating a significant normal stress difference, in gross contradiction of
the extended $\alpha$--viscosity prescription. These two facts are connected \citep{Will2004b}.

It should be clear by examining BNST95, HGB95, HGB96, as well as \cite{MatTaj95}, that the magnetic field that is being
produced by the MRI is being dragged by bulk shear of the fluid, as ideal MHD tells us it should, and it is statistically aligning itself with the
direction of mean shear. It is precisely by such dragging of field lines that the Maxwell stress in a turbulent medium produces a 
quasi-viscous off-diagonal stress, but why should the dragging of field lines stop there? Let us imagine, as I did in W01, a process where a turbulent or tangled field
is constantly being created, distorted by shear, and ultimately dissipated, as appears to be the case in these shearing-sheet simulations
(see fig.~2 in \cite{Will2004b}). Note that in equilibrium, when the turbulence
is saturated, this sequence is a logical sequence, not a chronological sequence. 

In fact, compare the magnetic stress $M_{ij}$ from eq.~(4) of W01 to the normalized stress $M_{ij}$ of eq.~(24) of BNST95, keeping in 
mind the switch $x\rightleftarrows y$ in going between the two papers, as well as the change of sign of $M_{xy}$ according to the differing
orientation of the shear. A least-squares minimization of the difference between each of the four nonzero members of the six independent stress components taken in turn
gives a fit of the normalized $M_{ij}$ of W01 to the $M_{ij}$ results of BNST95, written in the
coordinate system of the latter, which corresponds to pure azimuthal shear in a disk with ordering $(R,\theta,z)$, of
\setcounter{equation}{2}
\beq
%\begin{split}
(M_{ij})_{\rm W01}^{\rm fit}
% &= 
=
\left(
\begin{matrix}
0.02  &  -0.14  &  0 \cr
-0.14 &  0.96  &  0 \cr
0      &  0      &  0.02 \cr
\end{matrix}
\right)  
\tag{2a}
%\\
%%\nonumber
\eeq \beq
(M_{ij})_{\rm BNST95}^{\rm norm}
% &= 
=
\left(
\begin{matrix}
0.03  &  -0.09  &  0 \cr
-0.09 &  0.91  &  0 \cr
0      &  0      &  0.06 \cr
\end{matrix}
\right)
\tag{2b}
%\\ \end{split}
\eeq
for a value of the normed relaxation time $\gamma s = 6.9$ in the notation of W01.

In the case of the MRI the sequence of creation, distortion, and dissipation mentioned above is part of a larger feedback loop where the shear-aligned
magnetic field is unstable to the MRI (locally, through the combined action of the shear operator and the Coriolis operator), 
creating more turbulence and completing the feedback loop:
%\begin{figure}
%\beq
%\cdots \rightarrow M_{RR} \xrightarrow{\rm shear} M_{R\theta} \xrightarrow{\rm shear} M_{\theta\theta} \xrightarrow[{\rm (MRI)}]{\rm shear + rotation} M_{RR} \cdots
%\eeq
%\beq
%\xymatrix{
%U \ar@/_/[ddr]_y \ar@/^/[drr]^x
%  \ar@{.>}[dr]|-{(x,y)}             \\
%  & X \times_Z Y \ar[d]^q \ar[r]_p
%                 & X \ar[d]_f       \\
%  & Y \ar[r]^g   & Z                }
%\eeq
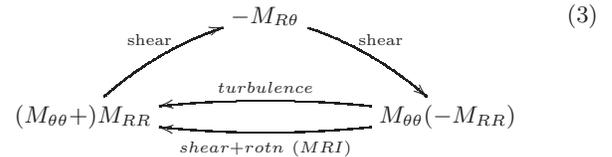
\beq
%\xymatrix{
%  & M_{R\theta} \ar@/^/[dr]^{\rm shear} & \\
%  M_{RR} \ar@/^/[ur]^{\rm shear} & & M_{\theta \theta} \ar@/^/[ll]^{shear + rotn\ (MRI)} \ar@/_/[ll]_{turbulence}
%}
\xymatrix{
  & -M_{R\theta} \ar@/^/[dr]^{\rm shear} & \\
  (M_{\theta\theta} + )M_{RR} \ar@/^/[ur]^{\rm shear} & & M_{\theta \theta} ( - M_{RR}) \ar@/^/[ll]^{shear + rotn\ (MRI)} \ar@/_/[ll]_{turbulence}
}
%\nonumber
\label{eq:graph}
\eeq
%\caption{The turbulent feedback loop as described in the text.}
%\end{figure}
In principle, the feedback loop may be closed by other sources of turbulence, such as convective or inertial
instabilities, or by MHD instabilities other than the MRI; it is in the nature of turbulence to stretch and contort material lines.
However, note that passive turbulent dynamo action (such as magnetoconvection) appears to be much less powerful than the active turbulent dynamo action of the
MRI (see HGB96 regarding this point).

Indeed, let it be clear that the process I have described here and previously (W01, W03, \cite{Will2004a} \cite{Will2004b}) depends upon dynamo
action in the sense of a turbulent dynamo as described by HGB96. Misunderstanding on my part regarding 
the definition of the word dynamo, a word which appears often to be taken to be synonymous with processes that can create self-sustaining
large-scale ordered fields, led me to state otherwise in \cite{Will2004a}, which was incorrect. On the other hand, the process does
not depend upon a mean-field dynamo.

In fact, a large-scale magnetic field will complicate this picture, as will other symmetry-breaking terms such as $\nabla P \times \vec \Omega$.
None of these are present in HGB96, nor are they present in the analysis here. A mean magnetic field is present in HGB95; the orientation
of this field seems to affect the effective relaxation time for the turbulence, as I note in W03; more significantly, the
presence of this field affects the saturation of the MRI. 
Ignoring the mean field and other symmetry-breaking terms for the present purposes however, 
I find that simple viscoelastic models of the stress produce a stress tensor $W$ that I write here
in terms of the thermodynamic pressure $P$ for a disk in which $v_R \ll v_{\theta}$ as (cf. Williams (2005) infra eq.~8)
\beq
{\bs W} / P = - \alpha_0 \;  {\bs I} - \alpha_1 \; \hat {\bs R} \hat {\bs \theta}  
+ \alpha_2 \; [\hat {\bs \theta} \hat {\bs \theta} - (1/3) {\bs I}]
\label{eq:elast}
\eeq
where  the set $\{\alpha_i\}$ consists of dimensionless constants, and $\hat {\bs R}$ and $\hat {\bs \theta}$ are unit vectors. 
That is, the stress is the sum of a turbulent
pressure equal to and parameterized by $\alpha_0$, a turbulent viscosity parameterized by $\alpha_1 \simeq \aSS$ (given that $P \simeq \rho c_s^2$), and a streamwise
normal stress difference,\footnote{
The first normal stress difference in the case of pure azimuthal shear is $N_1 \equiv W_{\theta \theta} - W_{RR}$. The second normal stress difference is 
$N_2 \equiv W_{\theta \theta} - W_{zz}$. I am only concerned with $N_1$ here; henceforth the first normal stress difference will simply be called 
the normal stress difference.}
\footnote{
Normal stress differences are not coordinate-invariant, but they are none the less useful and physical quantities within the context of the work here.}
 due to the turbulent elasticity, parameterized by $\alpha_2$. The form chosen for this final
term simply makes it traceless in 3D. For comparison with Williams (2005), $\alpha_0 = P_{\rm turb} / P$, $\alpha_1 = \mu_{\rm turb}(-R\partial_R\Omega) / P$,
and $\alpha_2 = 2 \zeta_{\rm turb} (R\partial_R\Omega)^2 / P$. For future reference observe that $\beta_p^{-1} = (3/2) \alpha_0 (-M_{kk} / W_{kk}) =
3 \alpha_0 ({\myM}_{kk} / W_{kk} ) \simeq (1.5  \ {\rm to}\ 2.3)\,\alpha_0 \simeq \alpha_0$. The lower number $(1.5)$ is derived from HGB96 and the higher number $(2.3)$ from
BNST95.

Note that the model for the MRI turbulent stresses given by \cite{Ogi2003} produces a different form
for the equilibrium turbulent stress, because of the explicit inclusion of the Coriolis effect 
through a Coriolis operator in his eq.~(23). This Coriolis operator rotates the Reynolds stress about the vector ${\bs \Omega}$
in the local corotating reference frame,
and it is most likely a crucial ingredient in a complete local dynamical model for MRI driven-turbulence.
However, this difference does not materially affect the conclusions here:
The existence of significant normal stress differences is an unavoidable feature of any realistic turbulence model for the MRI
in most, if not all, contexts, and this gross feature is present in the model of \cite{Ogi2003} as well.

Let me be the first to point out that there are three variables
to model (the three stress components), and I have done so using three parameters, so eq.~({\ref{eq:elast}})
not predict anything by itself, although note this prescription works quite well for describing the {full} six-component
stress tensor in 3D as well (W03). The reason this is a meaningful expression is that $\alpha_2 \simgeq \alpha_1$,
and that it makes physical sense to separate the normal stress into an isotropic and an anisotropic part.

\subsubsection{Magnitude of the normal stress difference and the effective Weissenberg-Deborah number}
The ratio of the normal stress difference to the viscous stress ($\alpha_2/\alpha_1$ in eq.~{\ref{eq:elast}}) is roughly proportional to the ratio
of the effective stress relaxation time $s$ to the shear time scale $(2A)^{-1}$, with the exact constant of proportionality being a model-dependent quantity.
For example, the Maxwell model of \cite{Ogi2001}, which I write here as
\beq
s(\hat {\bs {\cal D}}_t {\bs W}) 
+ {\bs W} = \nu [ \nabla {\bs v} + \nabla {\bs v}^{\rm T}] +\left(\nu_{\rm b} - {2 \over 3}\nu\right)(\nabla \cdot {\bs v}) {\bs I}
\eeq 
predicts $(W_{\theta\theta} - W_{RR})/W_{R\theta} = 2s(2A)$. 
(Note that $\hat {\bs {\cal D}}_t$ is the tensor generalization of the vector advection operator for the magnetic field ${\bs B}$, see \citet{Ogi2001}, W01, {\em et seq.}) 
As well, the simple $a$-$\delta$ model (model~B) of W03, namely
\beq
s(\hat {\bs {\cal D}}_t {\bs M}) + {\bs M} = a {\bs I}
\eeq
predicts $(W_{\theta\theta} - W_{RR})/W_{R\theta} = s(2A)$, under the approximation that 
\beq
W_{\theta \theta} - W_{RR} \simeq {\myM}_{\theta \theta} - {\myM}_{RR} = M_{\theta \theta} - M_{RR}.
\label{eq:approx}
\eeq
The $a$-$\delta$ model was not explicitly given in W01, although the construction in W01 gives a stress of exactly the same form, 
in the case of steady shear that is appropriate here.
The approximation in eq.~(\ref{eq:approx}) follows from the assumption
 that the turbulent Maxwell stress normal difference is much larger than the Reynolds stress normal difference,
as a result of either the quicker relaxation to isotropy of the Reynolds stress compared with the turbulent Maxwell stress,
or (more significantly) the assumption that the turbulent magnetic energy is greater than the turbulent kinetic energy.
This second assumption is important to the model of \cite{Ogi2001} as well; otherwise the use of the magnetic tensor advection operator for the
advection of the full turbulent stress would be unjustified.

Independent of the question of the value of the viscosity in a disk, then, is the question of the value of the ratio
of the normal stress difference to the viscous stress. Henceforth in this paper, I define the effective Weissenberg-Deborah\footnote{
Note that rheologists often distinguish between the Weissenberg number and the Deborah number; in principle it is possible to have
one large and the other small, see Section 4.2 of \cite{Phan-Thien}. Here I make no such distinction.
}
number $\We$ to be equal to the ratio $(W_{\theta \theta} - W_{RR})/|W_{R\theta}|$.
In the notation of
eq.~({\ref{eq:elast}) above and of \cite{Will2004b}, for reference, $\We = \alpha_2 / \alpha_1 = 2(\zeta / \mu)(2A)$.
It is also useful to define $\pi_0 \equiv \alpha_0 / \alpha_1$. The relative magnitude of the members of the set $\{\alpha_i\}$ are
then determined by $\pi_0$ and $\We$.

Basic dimensional arguments suggest that $\We$ is of the order of unity:
take the limit of an inviscid, perfectly conducting fluid, with zero mean field. Under the assumption that the  Coriolis 
parameter $\Omega$ and shear rate are comparable, the only local background dimensionful quantities available to construct a time scale
are $\Omega^{-1}$ and, given a length scale $L$ (such as a scale height), a sound-crossing time $L/c_s$. For a thin disk these two
are also both comparable, and this
 suggest that, barring some large or small dimensionless factor (such as the magnetic Prandtl number),
the effective relaxation time is comparable to the shear time scale (\citealt{Ogi2001}, W01, %\citealt{Will2001}, 
\citealt{Ogi2003}, W03, %\citealt{Will2003}, 
\citealt{Will2004b}).
One should then expect the streamwise normal stress difference to be comparable to the viscous stress, in gross contrast with the
extended $\alpha$-viscosity prescription, and this is what is seen. On the other hand, note that almost the exact same line
of reasoning has been used, historically, to argue that the original Shakura-Sunyaev $\aSS \simeq 1$, whereas the actual value of this parameter% may
is typically orders of magnitude smaller.
For reference, note that the two time scales above start out equal in the simulations of
HGB95; at the end of the simulations, $c_s / L \simeq 5 \Omega$ for the vertical field fiducial run
and $c_s / L \simeq 4 \Omega / 3$ for the azimuthal field fiducial run.

A much stronger argument can be made by appealing to the action of the MRI as a feedback loop, as in diagram~({\ref{eq:graph}}).
This feedback loop can only operate if the relaxation time of the process is long enough to create a substantial difference
between the streamwise
stress $M_{\theta\theta}$ and the cross-stream stress $M_{RR}$; otherwise the feedback loop will not close. This suggests that
$\We$ may be arbitrarily larger than unity but it is hard to see how it could be smaller than unity.

Even more significantly, an analysis of 
BNST95, HGB95, and HGB96 shows that %
that $\alpha_2$ is not just comparable to but in fact larger than both $\alpha_0$ and $\alpha_1$.
Some analysis of HGB95 is found in W03. For HGB96, the full stress tensor
(taking into account the often neglected
contribution of the magnetic pressure to the turbulent Maxwell stress, supra) from a naive 
average of their R1, R2, R3, R4, R6, and R7, is
$(W_{RR}, W_{\theta\theta},W_{zz},W_{R\theta}) = (-0.038, 0.012, -0.028, -0.015)$. The remaining two stress components, $W_{\theta z}$ and $W_{zR}$,
are not provided, although symmetry suggests they are zero.
I find $\alpha_0$, $\alpha_1$, and $\alpha_2$
to be $0.018$, $0.015$, and $0.046$, respectively; this gives $\We = 2.98$ and $\pi_0 = 1.19$.
The model eq.~(\ref{eq:elast}) is also an excellent fit to the full stress tensor provided in 
BNST95 as well. The full normalized stress tensor (taken from their eq.~(24), which provides all six components of the separate
Reynolds stress and the Faraday tension contribution to the turbulent Maxwell stress $M_{ij}$), again, correcting for the neglected contribution of the 
magnetic pressure to the turbulent Maxwell stress, and assuming a ratio of magnetic energy to kinetic energy of $6.67$ taken
from averaging over all runs given in their table~1, and normalized so that ${\rm Tr}(W) = -1$, is 
\beq
(W_{ij})_{\rm norm} = \left( %\matrix{
\begin{matrix}
   -.780 & -.159 & 0 \cr
   -.159 & .487 &  0 \cr
       0 &    0 & -.707 \cr
%}
\end{matrix} 
\right)
\eeq
under the coordinate ordering $(R,\theta,z)$ for the columns and rows.
This gives a value $\We = \alpha_2/\alpha_1 = 7.7$, larger than the value of $\We$ obtained from
HGB96. On the other hand, it is possible that the difference in $\We$ is simply reflective of the differing aspect ratio of these simulations.
As well, of course, future higher fidelity simulations may produce even different results.

Significantly, the variation in the value of $\We$ obtained in HGB96 compared with BNST95, 
{\em i.e.} roughly $8$ versus roughly $3$,
is much less than the variation between these two studies of the quoted value of $\aSS$, which differ by a factor of almost $10$.
This observation is in line with the observations of HGB95 [see their discussion in paragraph 1 of p.~749, and their eq.~(16) and eq.~(20)]
that, according to them, the traditional $\alpha$ parameterization `does not seem as appropriate here' because the viscous
stress correlates more strongly with the magnetic pressure than with the gas pressure. Since the field, and hence the magnetic pressure, is dominated by the streamwise
toroidal field, their observations combined with the observations in the paragraph above suggest that the ratio of the stress components,
and in particular, I would argue, the ratios $\alpha_0 : \alpha_1 : \alpha_2$, are a more robust result of shearing-sheet simulations than
the absolute value of the viscous stress as parameterized by the traditional $\aSS$.

\subsection{The form of the stress in a thick accretion flow}
\subsubsection{Steps towards a model}
In this section and the subsequent two sections, I describe basic steps towards a turbulence-based model of jets. A more complete description
of such a model will be given in a later paper. The stresses and forces to be considered here are the turbulent stress and the forces
due to gravity, the pressure gradient, and the centrifugal force. The centrifugal force is the result of the $\theta\theta$ component of the inertial stress tensor; the
remaining terms in this stress ($RR$, $Rz$ and $zz$) that appear in the meridional equation of motion
create a ram pressure which may create forces in the meridional plane that are of comparable magnitude to
the other forces described here and which may also aid in collimation. However, for the sake of simplicity here, these forces are ignored. 

\subsubsection{The fundamental assumption}

From the values given above for the set of numbers $\{\alpha_i\}$, in a Keplerian thin disk the resultant collimating force is even less than the radial thermal pressure gradient,
and thus negligible. Were this not the case, the thin-disk approximation would fail to be self-consistent.

Of much more interest is the case of a geometrically thick disk or flow. It is not yet clear how turbulence saturates in this context. 
A global accretion flow is different from a local shearing-sheet simulation in many respects. Among other things, the 
restrictive assumptions (quasi-periodic shearing box boundary conditions with no handedness) necessary 
to perform the shearing-sheet simulations of the MRI may affect the plasma $\beta_p$
at which the MRI saturates. In fact, as I discuss below, preliminary global simulations show the MRI to be more powerful (by which I mean $\beta_p$ is smaller)
than the local shearing-sheet simulations do. 

%For reasons that 
My approach here, as previously, is to fix the value of $\We$ (and $\pi_0$) for the reasons described above, but to let the
overall magnitude of the viscous stress be given by a free parameter. Ignorance of the behavior of turbulence in a thick disk
(or flow) is thus subsumed entirely into the viscosity prescription, and it is assumed that the other stresses (streamwise
normal stress difference and turbulent pressure) occur in proportion to the viscous stress.
Previously (W01) I fixed $\We$ to be of order unity.
If instead we take $\We \simeq {few}$ as above, then so much the better.

It remains to be seen what is the absolute value of the viscous stress $W_{R\theta}$ and how this compares with other dimensionally similar quantities.
In the next few sections I provide an unfortunate but necessary diversion into the various prescriptions for the viscosity.
Largely, the variations in ways of writing the viscosity are irrelevant to the gross conclusions.

\subsubsection{The form of the viscosity in W01 and W03}
The specific scenario discussed in W01 and W03 was a flow driven by the viscous spin-down torque on a central object
completely embedded in a thick accretion flow (I did not use the word flow but rather disk, but let it be clear that
this in no way should be taken to imply equilibrium as in the classic studies of thick equilibrium tori).
In W01 I chose to write $\nu$ in the form of a proportionality constant ($\alpha$) times a squared length scale times a rate. The rate I chose was
the shear rate at the fiducial radius of the central accretor $R_*$, so that $\nu \propto \left(2A\right)_* = q \Omega_*$.
 The length scale was somewhat idiosyncratically written in W01 as
$\sqrt{\beta} R_*$, and $\beta$ was a free parameter (not the plasma
beta) allowed to vary to take into account the uncertainty as to the correct length scale, and thus viscosity. (To avoid confusion
with the plasma $\beta_{\rm p}$, henceforth label $\beta$ from W01 as $\bW$;
also label $\alpha$ from W01 as $\aW$. Since $\aW$ is also
a free parameter and the two only appear in the form of the product $\aW \bW$,
 this second parameter $\bW$ is redundant in the notation of W01. Regrettably, this redundancy obfuscated the primary thrust of the paper.)
In particular, I allowed the length scale (here let me call it $L$) 
to vary between $R_*$ as a lower bound and the radius of the thick disk as an upper bound, which I denoted $r_m R_*$
(note the omitted minus sign in W01 and W03):
\beq
\begin{split}
%\nu = \aW \bW R_*^2 \left| {\partial \Omega \over \partial \ln R} \right|_*= \aW L^2 \left| {\partial \Omega \over \partial \ln R} \right|_*
\nu &= \aW \bW R_*^2 \left( {- \partial \Omega \over \partial \ln R} \right)_* \\ 
&= \aW L^2 \left( {- \partial \Omega \over \partial \ln R} \right)_* \\ 
&= \aW q L^2 \Omega_*
\label{eq:visc}
\end{split}
\eeq
The length scale $\sqrt{\bW}R_*$, or $L$, was assumed for simplicity to be a globally defined quantity that does not vary with position,
and so the viscosity above should be understood to be constant throughout the flow, for the sake of argument.

The reason for the use of the parameter $\bW$ was simply that it is sometimes taken to be implicit, when writing the viscosity in the form $\nu \sim \ell^2 \omega$,
 that the length scale $\ell$ and the rate $\omega$ are in some dynamical sense characteristic 
of the turbulence. I wished to
emphasize that, for a central object completely embedded in an accretion disk that is potentially much larger than the central object, this length scale could
be much larger than the radius of the central accretor, but no larger than some effective characteristic length scale of the thick accretion region.

This suggests, if $\aW$ in eq.~({\ref{eq:visc}}) is of the order unity, that the product $\aW \bW$ in principle may be much larger than unity.
Note that this does not necessarily imply anything about the value of $\aSS$, which may still be of the order of or much less than unity, see below.

\subsubsection{Shear versus vorticity versus Coriolis parameter}
The form I adopted above posits the viscosity being proportional to a shear.
\cite{AbBrLa96} performed shearing-sheet simulations of the MRI with variable effective $q$ and found that the viscosity
is proportional to the ratio of shear to vorticity, $\Omega / 2A$, or equivalently, $\nu \propto q/(2-q) c_s H$.
This is an important consideration to take into account for a model of the MRI for future work on thick disks and flows.
However in the rough work presented here this only introduces a factor of order unity which is negligible relative to the other uncertainties
and approximations in this work.

\subsubsection{Global versus local viscosity prescriptions}
Historically 
the turbulent viscosity has been variously parameterized in terms of purely local quantities, purely global quantities, and mixtures of
the two. For example, the Shakura-Sunyaev viscosity parameterization in its original form, supra, is a local definition, whereas the
viscosity parameterization of \cite{LBePri74}, namely $\nu \propto \Omega_* R_*^2$, is in global form.  Although \citet{LBePri74} do
not define an alpha, it is convenient to do so using their formulation, so let me here define
$\nu = \aLP \Omega_* R_*^2$. From this one may define an effective Reynolds number $\re = 1/\aLP$. 

Popular mixed forms for writing the viscosity prescription include $\nu = \alpha H^2 \Omega$ and  $\nu = \alpha c_s H$, where
$H$ is some measure of the disk thickness. 
These definitions are more or less equivalent to the original Shakura-Sunyaev prescription,
in the case of a thin disk. This redundancy disappears in the case of a thick disk or flow, however: it is not clear what is the
natural generalization of $H$ and $\Omega$ in a thick accretion flow, because there are are several nearly equal 
length scales and rates that are habitually conflated in thin-disk theory into these two quantities. 
For example, $H$ in thin disk theory is neither strictly local nor is it global, it lives somewhere in between. It is variously
defined to be a representative pressure scale height, a density scale height, the second moment of the vertical density distribution,
or the half-thickness of the optical depth $\tau = 2/3$ surface. It is not sufficient
to generalize it to a local quantity by defining it to be a pressure scale height, say, because strictly speaking the local
pressure scale height in a thin disk is infinite at the midplane. Similarly, $\Omega$ in thin disk theory is vaguely synonymous
with the orbital frequency, the Coriolis parameter, the Keplerian orbital frequency, the shear rate, and the vorticity; none of these
quantities are necessarily nearly equal one another in a thick disk.
This opens up the possibility
that $\alpha$ may be anomalously larger or smaller than the accepted value, depending upon what one means by $\alpha$, and
what part of a three-dimensional accretion flow one is examining.

The choice in W01 was to write the viscosity in a purely global form, similar to the Lynden-Bell--Pringle formulation above.
In particular, $\aLP = q \aW \bW$. Henceforth, I dispense with $\aW$ and $\bW$ in preference for $\aLP$.

The values of $\aSS$ and $\aLP$ are connected in a non-trivial way. Under the approximation that $P \thickapprox \rho c_s^2$ and
$2A \thickapprox \Omega$, one finds
\beq
{\aLP \over \aSS} \simeq { c_s^2 / \Omega \over \left(v_\theta^2\right)_* / \Omega_*}
\eeq
where $v_\theta = R\Omega$. Near the surface of the central accretor, then, ${\aLP / \aSS} \simeq c_s^2 / v_\theta^2$.
For $R \simgeq R_*$, one may have $\aLP \gg \aSS$ if $c_s^2 \gg v_\theta^2$.

Having now discussed the variations in writing the viscosity, I now return to the physics of collimation. 
Following W01, for concreteness, assume that in the thick disk, $\Omega \sim R^{-q}$, and in particular
$\Omega$ may differ greatly from the Keplerian value $\Omega_K$.

\subsection{The ordering of forces and speeds}
The force density that arise from taking the divergence of the stress tensor may be written as the sum of three
forces, see the three terms on the right hand side of eq.~(\ref{eq:tensordiv}) below, respectively called $F^{(1)}$, $F^{(2)}$, and $F^{(3)}$ here.
In this paper I do not consider  $F^{(3)}$.
 
The normal stress differences create a cylindrically radially inwards specific force $f^{(2)}_R ( = F_R^{(2)} / \rho ) = - (\We) W_{R\theta}/R = -(\We)\nu|\partial_R \Omega |$
(c.f. W01, eq.~6).
In the Shakura-Sunyaev local form, this may be written $f^{(2)}_R   = -(\We) \aSS c_s^2 / R$, and in the Lynden-Bell--Pringle global form,
$f^{(2)}_R = -(\We)\aLP R_*^2 \Omega_* (\Omega / R)$.
For ${\We} \simeq 1$ and for ${\re} \simleq 1$ (i.e. $\aLP \simgeq 1$) as assumed in W01, the collimating force due to the MHD turbulence is clearly significant. 

Consider here the relative strength of various specific forces. For simplicity here, assume that our test fluid element is close to the
equatorial plane, $z/R \ll 1$.
The set of forces I wish to consider here consists of the force due to the first normal stress difference $f^{(2) }_R$, 
the centrifugal force $f_c = \Omega^2 R$, and the force due to gravity $f_g = \Omega_K^2 R$.
The force due to the
turbulent pressure $f^{(1) }_R$ (which is essentially the same as the force due to the magnetic pressure 
because ${\myM}_{kk} \gg {R}_{kk}$),
and the force resulting from pressure $f_P = -(\partial_R P) / \rho$ will be discussed in the subsequent section below.

Before proceeding with the evaluation of forces, however, it is handy to describe the collimation stress
in terms of a wavespeed, just as \cite{AWML2003} make use of the Alfv\'en speed to describe the strength of the magnetic field.
A tangled field produced by a turbulent conducting medium supports Alfv\'en-like waves much like a viscoelastic
medium supports elastic waves (\cite{GruDia96}, \cite{Sch+2002}, \cite{Will2004b}). The speed of these transverse elastic waves $v_{\rm tew}$ is dependent upon
direction, for an anisotropic stress tensor. The wave speed due to the toroidal component of the tangled, turbulent field
is similar to the expression for the Alfv\'en wave speed, $v^2_{\rm tew}(\hat \theta) = \langle B_\theta B_\theta \rangle / 4 \pi \rho$. 
Then, for the toroidal mode,
\beq
v^2_{\rm tew}(\hat \theta) = {M_{\theta\theta} \over \rho} 
%\simeq {{\myM}_{\theta\theta} \over \rho}  \simeq { W_{\theta \theta} \over \rho} 
\eeq
Note that 
\beq
\begin{split}
M_{\theta\theta} &= \left( 2\alpha_0 + {2 \over 3} \alpha_2 \right) \, P \\
                 &= \left( 2\pi_0 + {2 \over 3} \We \right) \, \aSS \rho c_s^2 \\
                 &\simeq (\We) \aSS \rho c_s^2
\end{split}
\eeq
and therefore for the toroidal mode
\beq
v^2_{\rm tew}(\hat \theta) \simeq (\We) \aSS c_s^2.
\eeq
Typically the shearing-sheet simulations show $(\We)\aSS$ is small, on the order of $1/(few)$, e.g. $1/20$ for HGB96.
It is conceivable that high-fidelity global simulations will produce markedly different results; perhaps it is possible
that $(\We)\aSS$ becomes larger than unity. 
Incidentally, note that the principle axes of the stress tensor ${M}_{ij}$ do not
coincide with the coordinate axes, so an initially prograde or retrograde azimuthal wave will be refracted radially inwards or outwards respectively.
Henceforth let it be understood in this paper that by $v_{\rm tew}$ I mean the speed of the azimuthally traveling wave $v_{\rm tew}(\hat \theta)$.
Also note that $v^2_{\rm tew}$ is equal to the energy per unit mass stored in the azimuthal part of the turbulent magnetic field. The relative magnitude
of the various forces can be stated in terms of the relative magnitude of various speeds, or the relative specific energy densities as well, as given below.

The condition for the turbulent collimating force $f^{(2) }_R$ to dominate ({\em i.e.} have a larger magnitude than) the centrifugal force is
\beq
{v_\theta^2 \over c_s^2 } < (\We) \aSS,
\label{eq:centSS}
\eeq
or, equivalently,
\beq
\left( {R \over R_*} \right)^{2-q} < (\We) \aLP.
\eeq
Even more succinctly, eq.~(\ref{eq:centSS}) may be restated with the help of the wave speed expression above, so the condition becomes
\beq
v_\theta^2 < v_{\rm tew}^2.
\label{eq:centspeed}
\eeq
That is, the radial force due to the turbulent normal stress difference dominates the centrifugal force when the azimuthal flow velocity becomes sub-(quasi)-Alfv\'enic,
or equivalently, when the energy in the shear-aligned magnetic field is greater than the kinetic energy of rotation.
Interestingly, this critical field strength is the same as the MRI saturation field given in eq.~(8) of \cite{AWML2003}.

The condition for the turbulent collimating force $f^{(2) }_R$ to dominate gravity, in terms of the Keplerian orbital speed $v_K$, is
\beq
{v_K^2 \over c_s^2 } < (\We) \aSS,
\label{eq:gravSS}
\eeq
or, equivalently,
\beq
\left( {R \over R_*} \right)^{q-1} < (\We) \aLP \left[{ \Omega_* \over \left( \Omega_K \right)_* } \right]^2
\eeq
Again, more succinctly, eq.~(\ref{eq:gravSS}) may be restated with the help of the wave speed expression above, so the condition becomes
\beq
{G M_* \over R} = v_K^2 < v_{\rm tew}^2.
\label{eq:gravspeed}
\eeq
That is, the radial force due to the turbulent normal stress difference is stronger than gravity (but pushes in the same direction!) 
when the energy per unit mass stored in the turbulent azimuthal field exceeds the gravitational binding energy per unit mass.

Of course, there is nothing magical at all in eq.~(\ref{eq:centspeed}) and eq.~(\ref{eq:gravspeed}). The question is whether the turbulence
is ever so strong that these conditions are ever approached. Roughly, for collimation by turbulence to work, 
it appears likely that we must
have that $v_{\rm tew}^2$ is a substantial fraction of $v_\theta^2$, and preferably
$$v_\theta^2 < v_{\rm tew}^2,$$
 and then if we assume $(\We)\aSS < 1$ this implies
\beq
v_\theta^2 < v_{\rm tew}^2 < c_s^2.
\label{eq:speeds}
\eeq

The end result, perhaps not surprisingly, is that the turbulent elastic radial force $f^{(2) }$ is unequivocally important when the rotation
and the Keplerian orbital speed become sub-(quasi)-Alfv\'enic as in eq.~(\ref{eq:centspeed}), and furthermore subsonic as well
if we assume as is usual that $c_s^2 > v_{\rm tew}^2$. These are not  conditions that are ordinarily met in a thin disk,
but it is conceivable that they are met in portions of
 a thick disk, where radial pressure support becomes significant. Note that the Shakura-Sunyaev formulation
implies that turbulent pressure and hence magnetic pressure occurs
 in proportion to the thermal pressure. Even if this is not precisely true
in reality, it suggests that if we are to discuss radial thermal pressure support, then we should as well address radial magnetic
pressure support.

%I may place this in the language of our more recent work here. Assume $P/\rho \sim G M / R,$ i.e. $c_s^2 \sim R^2\Omega_K^2$.
%Note that $f_R^{(1)} \equiv F_R^{(1)}/\rho \sim \alpha ({\We}) P / (\rho R)$ Then for $\alpha \equiv W_{R\theta} / P$,
%$$
%f_R^{(1)} = \left( { \alpha ({\We}) \over \gamma} \right) {c_s^2 \over R}
%$$
%where $\gamma$ is the usual ratio of specific heats for a gas with a simple gamma-law equation of state.

\subsection{The role of magnetic and turbulent pressure}
This brings me now to address one final potential objection to collimation by turbulence-induced magnetic hoop-stresses, namely that
magnetic pressure might push out and counteract the inward pull of the hoop-stress.

This was explored in W01. As I describe in further detail here, the existence of magnetic pressure
may actually be beneficial to the jet-producing scenario I have described. It does, however,
lead to a focus on the normal stress differences, rather than the hoop-stresses themselves.

The phrases hoop-stress, azimuthal stress and toroidal stress are all succinct, equivalent and physically intuitive.
The phrase `positive normal stress difference between the azimuthal stress and the radial stress' is more precise,
but not succinct. 
One may have toroidal hoop-stresses without normal stress differences, but this is physically equivalent to a negative pressure.
Collimation in such a scenario can only occur through the radial gradient term below.

%Note that the turbulent stress tensor may be written
%\beq
%W_{ij} = \left< - \rho u'_i u'_j + B'_i B'_j / 4 \pi - (B')^2 \delta_{ij} / 8\pi \right>
%\eeq
%where primes denote fluctuating components of a quantity and we are purposely vague as to what type of averaging is implied by the angle brackets.
The existence of a positive (tension) turbulent magnetic hoop-stress requires a normal stress difference: the on-diagonal turbulent magnetic stress
can not be of the form of a negative pressure; the normal stresses can not all be positive, since their sum is negative semi-definite.
%The differing {\role}s of of these may be seen directly in the $R$-component of the divergence of a symmetrical (stress) tensor in cylindrical coordinates,
Recall that the $R$-component of the divergence of a symmetrical (stress) tensor in cylindrical coordinates is given by
\beq
\left( \nabla \cdot {\bs W} \right)_R = \partial_R W_{RR} - { W_{\theta \theta} - W_{RR} \over R} + \partial_z W_{Rz}.
\label{eq:tensordiv}
\eeq
For a purely azimuthal field, spatially stochastic or otherwise, the stress $W_{RR}$ is negative and due entirely to magnetic pressure, and it appears in the force equation
through its gradient. In contrast, the normal stress difference $W_{\theta\theta} - W_{RR}$ is due entirely to the magnetic Faraday term, {\em i.e.} the tension
of field lines, and it appears in the force equation only through the geometrical factor $1/R$, not through its gradient.
In the more general case of a tangled field which is predominantly but not entirely azimuthal, not only will $W_{\theta\theta}$ have some contributions
from the magnetic pressure, but $W_{RR}$ will have some positive contributions from the tension
term as well. However, the normal stress difference  $W_{\theta\theta} - W_{RR}$, absent Reynolds stresses, is still due entirely to the magnetic tension term. 

Now let us examine outflow collimation in engineering. Consider jets from nozzles, such as
a garden hose nozzle or a rocket nozzle. These jets are confined and collimated at their base by the anisotropic elastic stress in the
collimating medium ({\em i.e.} the nozzle). Similarly, the explosive gases that propel the slug of a rifle or cannon are effective at propelling the projectile
becuase of the anisotropic elastic stress in the surrounding barrel. The metal in the 
barrel is radially in compression (negative stress) and azimuthally in tension (positive stress).
In the case of both the steady dynamics of nozzles and the
transient dynamics of guns, the collimation is thus due to a
normal stress difference in the surrounding material. In all cases, the anisotropic stress of the surrounding medium reshapes the effect of an
isotropic stress (namely the pressure) to create an anisotropic outflow.

On the other hand, it is well to note that underwater detonations of explosives may produce jet-like columns of water; as well, large thermonuclear
explosions in the stratified atmosphere of the Earth can cause a rapid vertical acceleration of material. An analogous phenomenon is supernovae-fueled
outflow in stratified galactic atmospheres, known as the galactic fountain. Such processes create anisotropic outflows
 due to collimation by pure pressure.
 
The magnitude of the force density due to the gradient of turbulent pressure (including magnetic pressure)
$F^{(1)}_R = \partial_R (W_{R R})$ should be
of the order of $W_{R R} / R$. It remains to be seen whether this force is collimating or
anti-collimating.

I assumed in W01 that pressure increases as we near the central object, so that the
pressure is anti-collimating. Then, $F^{(1)}$ and $F^{(2)}$ act oppositely. For the net force to be collimating,
one requires $|F^{2}| > |F^{(1)}|$.
I argued that work done by the magnetic curvature forces -- {\em i.e.} work done by the collimating force
$F^{(2)}$ -- is work done against the magnetic pressure gradient as material is advected inwards near the equatorial plane.
In other words it acts to increase the value of the Bernoulli $b$ parameter, potentially changing the sign of $b$ from negative to
 positive${}^6$.
This work is then
available to power an outflow  (c.f. W01, eq.~8).

Further study and simulation should help elucidate whether turbulent magnetic pressure is indeed anti-collimating or
collimating. It is possible the answer depends upon the location in the flow being examined.

\section{COMPARISON WITH COLLIMATION BASED ON DISPERSION RELATION}
\subsection{Introduction}
As we shall see below, the estimates for the strength of the collimation as discussed above and in W01 and W03
are consistent with the estimates for the strength of the MRI argued by \citet{AWML2003}.

It is well known that fluid dynamical instabilities that lead to turbulence may be self-limiting,
in the sense that the turbulence creates transport that, in many cases, damps out the driving terms in the instability.
Thus, for example, an adverse entropy gradient in a stellar interior drives convection, which in turn reduces the adverse entropy gradient.
Similarly, it was suggested by \citet{AWML2003} (see also \citealt{AkiWhe2002})
that, under conditions in a core-collapse supernova where the MRI is
locally unstable, a small seed field should grow via the MRI until the field reaches the critical field strength at which
the MRI is no longer unstable. Since the instability grows exponentially, it was argued that it dominates linear-in-time processes such as passive
field wrapping.
The authors found that given initial angular velocity of the supernova progenitor core of the order of a few radians per second, the combination of collapse and
conservation of angular momentum on shells combine to create strong MRI-unstable differential rotation in core collapse.
Further they point out that, although the ultimate field orientation is not clear, a toroidal field configuration
seems likely, considering the large differential shear in core collapse.

%In the simplified case of negligible entropy gradient, the instability criterion reduces to (Balbus \& Hawley 1991, 1998)
%\beq
%{d \Omega^2 \over d \ln R} + \left( \bf k \cdot \bf v_{\rm A} \right)^2 < 0.
%\eeq
%AW02 and 
\subsection{Estimates for the field saturation}
\citet{AWML2003} provide a number of different estimates for the saturation field of the MRI.
One estimate may be obtained by setting the
length scale $\ell \sim dR / d(\ln \Omega)$ equal to the characteristic mode scale. This results in a field strength
$%\beq
B^2_{\rm sat} \simeq 4 \pi \rho R^2 \Omega^2
$%\eeq
 or, equivalently, $\left(v_{\rm A}\right)_{\rm sat}^2 = v_\theta^2$.
\citet{AWML2003} point out that this estimate of the field strength is substantiated by the simulations of HGB96,
namely $2 \pi B_{\rm sim} = B_{\rm sat}$. In point of fact, the results of HGB96 refer to the mean square of the
field components, as the mean field in these simulations is identically zero, as I discuss at length above.
From the point of view of calculating the stresses and forces due to the field, the distinction is largely irrelevant.
This distinction does make a large difference, however, when it is argued that the MRI-induced stresses of \cite{AWML2003}
depend upon processes such as helicity conservation that, through inverse cascades, create a significant mean field.
While such processes may indeed occur in core-collapse supernovae, I would argue that \cite{AWML2003} may be re-read
with an understanding that the mean field may be replaced throughout their analysis with a tangled field with a mean stress,
and this difference in interpretation does not change the analysis at all.

Another estimate is related to this first estimate by setting the mode length equal to the local radius, resulting
in a field strength
\beq
B^2_{\rm rad} \simeq 4 \pi \rho R^2 \Omega (R \Omega' / \Omega)^2
\eeq

Other estimates are found using the maximum unstable growing mode (with a vertical seed field), or by setting the wavelength
of the maximum growing mode equal to the shear length.

Again, one of the issues that is naturally raised in their analysis as well as in my analysis is the question of local time scales
and length scales. For example, the difference between the estimates $B_{\rm rad}^2$ and $B_{\rm sat}^2$ is a factor of the
ratio of the shear to Coriolis parameters.

Finally, they find the ordering $$c_s \gg r\Omega \sim v_{\rm A}.$$ In particular, they find that the plasma $\beta_p$ is large,
specifically the magnetic pressure is on the order of a few \% of the gas pressure. They none the less argue that the resultant
MHD luminosity may be significant enough to propel and collimate jets in core-collapse supernovae.

Note that the similarity in the conclusions regarding the magnitude of the hoop-stresses based on turbulence modeling and
on the dispersion relation shows how the  
 the saturation level of the MRI and the value of the effective viscosity is related to the MRI dispersion relation.

\cite{AWML2003} do not address the magnetic pressure in their work.

%\subsection{Direct Comparison with My Work}

\section{COMPARISON WITH SIMULATION}
Let me now briefly address recent simulations in the light of the preceding discussion.

Only lately has it become possible to perform global DNS magnetohydrodynamic accretion flow simulations in three dimensions that are capable of
showing the action of the MRI. These global simulations (Hawley, Balbus \& Stone 2001) show, as the authors point out,
that the behavior of the flow is not adequately described by a hydrodynamic $\alpha$-model. 
This should not be surprising because, 
by inspection as I discussed above and previously, neither are local simulations adequately described by a hydrodynamic $\alpha$-model.

Hawley, Balbus \& Stone (2001) simulated the non-radiative
 magnetohydrodynamics of an initially constant angular momentum torus in three dimensions. 
They gave meridional contour plots of the magnetic pressure and the off-diagonal ($R-\theta$) Maxwell stress, but as these
plots are unscaled and as they do not show the other stress components, it is impossible to see if the MRI is creating
hoop stresses through the tension term in the magnetic stress. The magnetic pressure appears to be anti-collimating
(as I argued in W01) in the equatorial plane, and collimating above the plane. 
They found that after several %about four or five 
orbits the MRI creates magnetic field with plasma $\beta_p$
on the order of roughly $5$--$10$ close to the equatorial plane, whereas several scale heights above the equatorial
plane $\beta_p$ is less than $1$. Thus, even on the equatorial plane the MRI saturates to a level much higher ({\em i.e.}~$\beta_p$ is
smaller) than in the zero mean-field shearing-sheet simulations of HGB96, which yield a plasma $\beta_p$
of roughly $40$, and the MRI saturates rather to a level comparable to the nonzero mean-field simulations of HGB95, which yield
a plasma $\beta_p$ of the order of $7$ for the azimuthal field runs and of the order of $2.4$ for the vertical seed field (as determined
from a na\"ive average of all the runs on their tables~1 and 3, and excluding run Z3). Note that in the case of HGB95 these values for the $\beta_p$ include contributions
due to the mean field and the turbulent field.

The further exploration of a very similar set of simulations in Hawley \& Balbus (2002) allows us to place these simulations more firmly in the context
of our remarks in the previous sections. This second set of simulations has been applied to the very hot ($\sim10^{13}\,{\rm K}$)
accretion flow of Sgr~A*; nevertheless the authors use a very simple single-component equation of state
which allows one to estimate the ratio of sound speed to orbital speed from their fig.~8, namely $$c_s^2 \simeq (1/10)v^2_\theta.$$ 
Using $\beta_p \simeq 5$ for the saturation of the MRI on the equatorial plane, then, I have $$v_{\rm A}^2 \simeq (1/250)v^2_\theta,$$
and $$v_{\rm A}^2 < c_s^2 < v_\theta^2.$$ On the other hand, the situation changes above the equatorial plane where $\beta_p^{-1}$ is
much larger, with a correspondingly larger Alfv\'en speed relative to sound speed.
Here, $v_A^2$ is likely a much larger fraction of $v_\theta^2$ or $v_{\rm K}^2$, but it is difficult to tell by how much.

Collimation by the MRI appears to be seen in the $2.5{\rm D}$ simulations of \cite{Kud2002}, \cite{ProBeg2003},
and \cite{PMAB2003}. Note that all three components of the velocity and
magnetic field are included in these simulations. However, the constraint of axisymmetry in these simulations
necessarily restricts the action of the MRI; in particular, this constraint eliminates the non-axisymmetric modes which act upon the azimuthal field, and which I
called upon to complete the feedback loop in the creation of turbulent azimuthal hoop-stresses in equilibrium that were the
basis of my analyses. There therefore remains the question of the extent to which the collimation in \cite{Kud2002},
\cite{ProBeg2003} and \cite{PMAB2003} is a transient effect. This may be a moot point in the case of collapsars, but it is essential to the
case of steady accretion.% onto a black hole.

\cite{ProBeg2003} present
axisymmetric simulations of the accretion of low angular momentum gas onto a black hole. Despite the initial low angular momentum,
at late stages of the accretion the Alfv\'en speed, sound speed, azimuthal fluid velocity and Keplerian velocity are all roughly comparable
(see their figures~9 and 10).  In contrast, in the collapsar simulations of \cite{PMAB2003}, these velocities sometimes differ
by more than an order of magnitude, and again $v_{\rm A}^2 < c_s^2 < v_\theta^2$.
In both cases, the radial pressure gradient in the equatorial plane is anti-collimating.
Also in both cases, the field is clearly dominated by the toroidal component.

\cite{PMAB2003} state that it is the gradient of this toroidal field in particular that drives the outflow.
Let me point out that in the meridional plane, the gradient of the toroidal field appears only
through the magnetic pressure term. By rough examination of the printed results of Proga~et~al.~(2003), it appears the
toroidal field exerts a force through the curvature term
that is comparable to the force due to its gradient. 
That is, I argue
that the term $B_{\phi}^2/r$ in the divergence of the stress may be of approximately equal significance to the results of
\cite{PMAB2003} as the gradient term $\partial_r (B_\phi^2)$ that the authors address. (As an aside, note that also of comparable importance,
depending upon location in the flow, is, by inspection, the ram pressure from infall, so that it appears that the jet is in part not only
magnetically confined but inertially confined as well.)

A final intriguing note regarding the simulations of \cite{PMAB2003} is that the toroidal field component changes polarity with time and location.
This confirms that the MRI can collimate jets through a disorganized rather than an
organized (mean-field) toroidal field, as I suggested in W01 and W03. Note that the possibility of jet collimation by a tangled field has also been
investigated in the context of the far-downstream jet by \cite{Li2002} 
who showed that %, in steady jets such as in AGN, 
a tangled field avoids the long-mode kink instabilities that plague jets collimated by an ordered field.

Finally, let me address the recent core-collapse simulations of \cite{TQB} who insert MRI-induced viscous dissipation into 1D simulations.
\cite{TQB} point out that \cite{AWML2003} overestimate the viscous time scale, and that viscous process that smooth out gradients in $\Omega$
must be taken into consideration on the time scale case of core collapse. 
As a result, viscous heating may significantly alter the core collapse dynamics.
The proof-of-concept work of \cite{AWML2003}, however, demonstrates that one
must take into account the streamwise hoop stress as well. 
My primary goal here and elsewhere has been
to focus upon the hoop stress, as this had not been addressed in the 
literature. However, a full accounting must at the very least address all three components, as I have
attempted here and elsewhere,  and not just one or another component. (The physical significance of the remaining three stress components, which may come
into play when one moves above or below the equatorial plane of symmetry, is not explored here.) For example, the viscous stress (and not just the normal
stresses) is important to the model of W01, as this stress transports energy and angular momentum, which the other stresses do not.
 My analysis of the results of
\cite{TQB} shows that at, say, the fiducial radius of $20{\rm km}$, the inwards MRI-induced force in the equatorial plane, which the authors do not include in their code, 
is no more than half an order of magnitude less than the outwards centrifugal
force that the authors do include in their code, if one takes the modest value $\We = 3$.
Admittedly, both sets of simulations are severely limited by the inherent assumption of spherical symmetry of a 1D code.
These are valuable proof-of-principle studies; % that use 1D simulations
future progress demands multidimensional MHD codes.

\section{CONCLUSION}
In this paper I have discussed my previous work (W01, W03, Williams 2004a) regarding the hypothetical creation and collimation of jets by
MHD turbulence and by the MRI in particular. I have added on to this work with a much more extensive and in-depth discussion of the dynamics
of turbulent stress and the modeling of turbulent stresses in accretion flows. Also, I have placed this work within the context of other
recent work, in particular the semianalyitic work of \cite{AWML2003} based on the dispersion relation for the MRI,
as well as the DNS simulations of \cite{HBS2001}, \cite{HawBal2002}, \cite{ProBeg2003}, and \cite{PMAB2003}.

I have focused in this paper on expounding on how simple MHD turbulence modeling considerations in the context of steady shear,
as discussed in \cite{Will2004b}, lead to the conclusion that MHD turbulence and MRI-driven turbulence in particular can create
collimating forces. As discussed previously (e.g. W03) I also show how collimation can be seen directly in the results of
local shearing-sheet simulations of the MRI, which significantly predate global MRI simulations. 
Turbulent collimation by quasi-viscoelastic stresses, as discussed, has a natural
analog in the Weissenberg effect, whereby viscoelastic fluids climb a spinning rod. Thus, the turbulent MHD collimation of astrophysical
jets as discussed has a natural analog to the prehaps more prosaic physics of the dynamics of ordinary viscoelastic fluids such as egg whites
which climb spinning rods.

Indeed, this research was originally inspired by my serendipitous discovery in the literature of a laboratory viscoelastic phenomenon
well-known to rheologists (\citet{Gie1963, ThoWal1964, NCFMF}),
namely the viscoelastic flows discussed in W01,  which
seemed to bear a remarkable similarity to jet phenomena. This lead naturally to the question of whether turbulent conductive fluids supporting
a tangled magnetic field can in some ways behave like viscoelastic fluids, and thus create a collimating force by tangled fields in
analogy to the collimating force due to normal stress difference in polymeric laboratory viscoelastic fluids. Fortuitously
enough for this line of reasoning, simulations of
the MRI in particular demonstrate that MRI-induced turbulent stresses do in fact look in many ways like the stresses in viscoelastic, as opposed
to viscous, fluids.

Of course, turbulent magnetized fluids are not really viscoelastic fluids, and there are many fundamental
questions that remain regarding the behavior of MHD turbulence in accretion flows. Here I discuss four in particular:

\vskip 5mm

The primary open question is not whether the MRI can create confining magnetic hoop-stresses,
but rather whether these stresses are strong enough to be a dominant confining mechanism for jet creation. 
Indeed, the ordering of speeds given in eq.~({\ref{eq:speeds}}) may be difficult to acheive in steady-state accretion flow scenarios.
The weaker condition that the turbulent magnetic fields be dynamically significant, but perhaps not strong enough by themselves alone
to create an anisotropic outflow, can be restated that the ratio $v_\theta^2 / v_{\rm tew}^2$ not be much larger than unity.
Conventional wisdom would suggest that this condition is difficult to meet in accretion flows in AGN and protostellar systems.
On the other hand, our understanding of the physics of these systems is far from complete. Below, I offer as a hypothesis (see hypothesis
(1)) the notion
that in these accretion systems, when jets are produced, $v_{\rm tew}^2$ is a substantial fraction of, if not larger than, $v_\theta^2$, at the
jet base.

Regardless of the validity of this hypothesis,
it does appear that, as I argued in W01 and W03, the MHD turbulent hoop-stresses in accretion are at least of equal magnitude to the
turbulent viscous stresses. Thus, global simulations of the inner regions of  accretion and jet launching that treat
turbulence with a purely viscous $\alpha$ model are of questionable physical relevance, as pointed out by
Hawley~et~al.~(2001). For example, this result casts doubt on the results of, {\em e.g.}, \cite{AIQN2002}, because these
authors do not include the turbulent hoop-stresses in their radial force balance equation, and as noted above, the resultant
force changes the value of the Bernoulli $b$, which must be taken into consideration.
Furthermore, there is a simple and computationally cost-effective way to include these hoop-stresses
in 2D, by the use of a viscoelastic turbulence model rather than a viscous model.
% or even better yet, a turbulence model
%that incorporates the relevant instabilities dynamically, such as the model of \citet{Ogi2003}.

A second question,
from the point of view of turbulence modeling, is the relative magnitude of
 of the MRI versus other instabilities
(such as convection and, possibly, inertial instabilities) in creating turbulent hoop-stresses. As has been argued previously,
the MRI, as a magnetohydrodynamic instability rather than a purely hydrodynamic instability, injects energy directly into
the Maxwell stress tensor rather than indirectly, such as the case in convection, where energy may be expected to be transferred
to the turbulent Maxwell stress by coupling to the Reynolds stress. Thus it is quite reasonable to expect that turbulent Maxwell stresses
are relatively more significant than Reynolds stresses in purely MRI-driven turbulence, and this appears to be the case. In fact in W01 I assumed
that the turbulent Maxwell stresses dominate the Reynolds stresses. This is not a necessary condition for the creation of collimating
hoop stresses, but it helps, because otherwise one must contrive that the relaxation time for the Maxwell stresses is much longer than
the relaxation time for the Reynolds stresses, since the action of the shear operator on the former creates positive (tension) streamwise
stress whereas the action of the shear operator on the latter creates negative streamwise stress. In addition,
in the case of the MRI the dimensional arguments presented above and in W01 and W03 argue that the effective relaxation time
is roughly equal to $\Omega$ or $A$. These arguments no longer hold when shear and Coriolis forces no longer drive the instability.

A third question is the relative importance of tangled versus organized fields in this collimation. 
It remains to be seen the extent to which mean-field dynamo processes
may be significant in MRI-generated collimating
forces.
Rather, here and previously
I have emphasized tangled fields. I argue collimation by MRI-induced tangled fields to be a more robust result than collimation by MRI-induced organized
fields. First, I have argued above that simulations may be interpreted as showing that the creation of small-scale fields is a more reliable
feature of the MRI than creation of large-scale mean fields. Second, if the MRI fails to create a small-scale tangled field, then it also fails
to provide a mechanism for turbulent transport of angular momentum. This would quite clearly be a problem since
 it is currently accepted that turbulence driven
by the MRI is most likely largely if not solely responsible for angular momentum transport in ionized disks. A third point, particularly 
apropos in the case of transient accretion events such as collapsars, is that the creation of a tangled field may proceed faster than
the creation of a large-scale mean field, to the extent that creation of a mean field depends upon inverse cascades acting over several orders of
magnitude in length-scale.

Thus, \cite{LyPaBl03}, in the context of gamma-ray burst (GRB) fireballs, argue that the prompt creation of a large-scale
field is problematic. However, citing \cite{GruWax99} who in turn cite \cite{LoePer98}, they argue that the linear polarization
of GRB0212206 seen by the {\em RHESSI} satellite as described by \cite{CobBog2003}
implies a large-scale
field [as indeed \cite{CobBog2003} argue as well]
 and that, since it is difficult to create such a field quickly, the field must exist prior to the catastrophe
producing the GRB:
`Such fields cannot be generated in a hydrodynamically-dominated outflow, which is causally
disconnected on large scales. Thus, the large scale magnetic fields should be present [at the beginning].', \cite{LyPaBl03}.
 The authors claim that `[A large polarization] implies that magnetic
field coherence scale is larger than the size of the visible emitting region\dots'. The truth of this assertion depends upon
 what is meant by `coherence scale'.
In particular, it is wrong to assert that the polarization implies a large-scale field:
tangledness should in no way be confused with isotropy.

As an analogy, the orientation of mesogens in the nematic phase of liquid-crystal displays (LCDs) is coherent
on a macroscopic scale. This large-scale anisotropy (or {\ae}olotropy}) makes the speed of light depend upon polarization,
even though the size of an individual
mesogen is typically smaller than an optical wavelength (Rosenblatt, private communication).
Likewise, a large-scale mean dyad field $B_iB_j$ does not imply a large-scale mean vector field $B_i$.
Processes that do not discriminate between ${\bs B}$ and $-{\bs B}$ care about the former, not the latter;
this includes the linear polarized emission of cyclotron and synchrotron
radiation.
In the case of cyclotron and synchrotron radiation, only the presence of circular polarization discriminates between a mean field and
a mean dyad (\citet{matio2003}). Thus, large linear synchrotron polarization is possible due to a tangled field, so long as $M_{ij}$ is
strongly anisotropic (\citet{sagiv2004}).

Again regarding tangled versus ordered fields, \cite{TQB} state, regarding the work of \cite{AWML2003}, that `[The argument presented
by \cite{AWML2003}] assumes that the magnetic field generated by the MRI can form
the organized large-scale fields required for collimation and jet formation.'
Strictly speaking this may be true, since \cite{AWML2003} refer to mean fields rather than tangled fields.
But, inasmuch as \cite{AWML2003} may be reinterpreted as describing the magnitude of the stress due to a tangled field
rather then an ordered field as described above, the estimates for the forces due to the MRI remain the same. Thus, while the authors of \cite{AWML2003}
were perhaps unaware of it, the processes they described does not actually depend upon
the creation of organized large-scale fields, and in this sense \cite{TQB} are incorrect.

Fourth and finally, the {\role} of magnetic turbulent pressure and other stresses such as the neglected inertial stresses requires further analysis.
I have argued here that the turbulent magnetic pressure will be anti-collimating in jet scenarios, and that it provides a reservoir of energy
to accelerate the jet, but this conclusion is more tenuous than the more robust conclusions regarding the collimating normal stress differences
caused by MHD turbulence.
In addition, is conceivable that neither magnetic stresses nor inertial stresses are large enough by themselves to launch and collimate
jets, but that the two acting together may create jets. Discussion of intertial stresses in particular is left to future work; notwithstanding
their potential dynamical significance, I hypothesise below that the MHD turbulent stresses must be dynamically significant in a broad range
of jet-producing accretion flows.

These problems discussed above, as well as many others, remain to be answered completely. None the less,
I have shown here that my previous estimates of the turbulent collimating force
are consistent with other estimates, and in particular the three lines of evidence and reasoning I presented above
all reduce to the same condition, which may be stated in two different ways: 
The energy in the toroidal field is a sizeable fraction the orbital kinetic energy, or equivalently, 
the local toroidal magnetic wave speed is a sizeable fraction of the orbital speed.
The difference between the theoretical approach based on the MRI dispersion relation and the approach based on turbulence modeling is simply whether
the wave in question is a purely Alfv\'en wave or an Alfv\'en-like turbulent transverse elastic wave.
It is not yet clear how sizeable a field is necessary, and whether condition (20) need indeed be met anywhere in the flow for a
turbulent toroidal magnetic field to be dynamically significant.  

Notwithstanding these remaining questions, then, I offer as hypotheses the notions that:
\vskip 5 mm

(i) MHD turbulence generated by the MRI
(as well as perhaps additional instabilities) creates dynamically significant toroidal hoop stresses (more properly, a toroidal normal stress
difference) that contribute significantly to jet creation and collimation; 

(ii) this process does not depend upon mean-field dynamo
processes, and in particular much of the field energy may be in a tangled field rather than a mean field;

(iii) inasmuch as
MHD turbulence (driven by the MRI in particular) is thought to be the predominant source of turbulent viscous angular momentum transport in hot
ionized disks and accretion flows, the corresponding turbulent elastic collimation process described here may
be a nearly universal aspect of jet formation, applicable not just to the transient dynamics of
 collapsars but to jets in a broad range of quasi-steady-state systems such as AGN and quasars, microquasars, and protostellar systems;
this hypothesis is offered notwithstanding the predominant understanding of thick accretion flows that would argue that such azimuthal turbulent
magnetic fields should be relatively weak;

(iv) that this jet creation and collimation, as described, is not offered as a process that occurs in addition to traditional magnetocentrifugal
acceleration mechanisms, but rather as an alternative to them. 
\vskip 5 mm

\cite{blapay1982}, \cite{shuetal1994} and others have argued that jets are created by material flung out following large-scale
open magnetic field lines. To be quite explicit, as a hypothesis, I %categorically 
reject the notion that such magnetocentrifugal mechanisms are responsible for tightly collimated jets.
In such scenarios, power can not be transferred to the jet without
transferring angular momentum as well. Note that recent observations of certain jets suggest that the fields may have
helical structure (\citet{helical}) and the jets may be rotating (\citet{rotate}). 
However, this in itself does not demonstrate magnetocentrifugal acceleration. 
Indeed, it would be surprising if Nature conspired to remove all of the residual angular momentum from material before accelerating
it in a jet. The presence of a helical magnetic field, rather than a purely azimuthal or vertical field,
 can  be explained as the result of advection, and does not necessarily
 say what is  the dynamical {\role} of that field, and in particular it does not necessarily imply that material is
in some sense being flung out along field lines.
Rather it is suggested here that the predominant {\role} of the magnetic
field in the central engine of a broad array of jet sources is the creation of turbulent hoop-stresses as described, which create and 
collimate an outflow deep in the core of the central engine. In analogy to viscoelastic laboratory phenomena, and in contrast to
magnetocentrifugal mechanisms, in such a scenario material
can take part in a collimated outflow not because angular momentum is transferred to it,
 but because angular momentum is removed from it. % and transported to the surrounding material.

\section{ACKNOWLEGEMENTS}
The author thanks J. C. Wheeler and R. A. Meisner for encouragement, S. Akiyama for discussions, and the anonymous referee. This
work was supported by the E.D.D. of the State of California.

\newcommand{\apj}{{ApJ}}
\newcommand{\ana}{{A\&A}}
\newcommand{\mnras}{{M.N.R.A.S.}}

\end{document}